\documentclass[showpacs,preprint,preprintnumbers,amsmath,amssymb,nofootinbib]{revtex4}

\usepackage{graphicx,bm}
\usepackage{dcolumn}    

\begin{document}
%

\title{\bf
Naked singularity resolution in cylindrical collapse
}

\author{Yasunari Kurita$^{1,2}$} 
\email{kurita_at_sci.osaka-cu.ac.jp}

\author{Ken-ichi Nakao$^1$} 
\email{knakao_at_sci.osaka-cu.ac.jp}  

\affiliation{ 
   $^1$Department of Physics, 
   Osaka City  University, Osaka 558-8585, Japan\\
   $^2$Yukawa Institute for Theoretical Physics, Kyoto University, 
   Kyoto, 606-8502, Japan
}

\begin{abstract}

In this paper, we study the gravitational collapse of null dust in the 
cylindrically symmetric spacetime. The naked singularity necessarily 
forms at the symmetry axis. We consider the situation in which null dust 
is emitted again from the naked singularity formed by the collapsed null dust 
and investigate the back-reaction by this emission for the 
naked singularity. We show a very peculiar but physically important case 
in which the same amount of null dust as that of the collapsed one 
is emitted from the naked singularity as soon as the ingoing null dust hits 
the symmetry axis and forms the naked singularity. 
In this case, although this naked singularity satisfies the strong curvature 
condition by Kr\'{o}lak (limiting focusing condition), geodesics which hit the 
singularity can be extended uniquely across the singularity. 
Therefore we may say that the collapsing null dust passes through the singularity 
formed by itself and then leaves for 
infinity. Finally the singularity completely disappears and the flat spacetime 
remains. 
\end{abstract}


\preprint{ OCU-PHYS-236 }
\preprint{ AP-GR-28 }
\preprint{ YITP-06-03}

\maketitle

\section{Introduction}
\label{sec:intro}

It is important to investigate the gravitational collapse in general relativity.
One of the most important motivations is to examine the cosmic censorship 
hypothesis \cite{Penrose:1969}. 
However, previous theoretical studies have implied the existence of counter 
examples to the hypothesis \cite{Joshi:1993}-\cite{Shapiro:1991a}.
If the universe is not censored, we can not deny the possibility that a 
spacetime singularity is observable. The observable singularity is 
called the naked singularity. By its definition, there are regions which 
suffer the physical effects of naked singularities. Therefore, in order 
to construct a  solution describing a spacetime accompanied by 
naked singularities by solving Einstein equations as a Cauchy 
problem, we need to specify the boundary condition at the naked  
singularity. In other words, in order to predict the dynamics of 
the spacetime accompanied by the naked singularity, we need the 
knowledge about the spacetime singularity. However all known theories 
of physics including general relativity are not applicable 
at the spacetime singularity. This fact is not a serious problem. Conversely  
the appearance of naked singularities might give us important 
information about new physics \cite{Harada:2004mv}. 
The knowledge of the naked singularity 
is very important even within the framework of general relativity. 

It is well known that, in the case of the cylindrical gravitational 
collapse, a resultant singularity is always 
naked \cite{Thorne:1972,Hayward:2000}. 
In the literature, there are a few exact solutions for cylindrical 
gravitational collapse. One of them gives a description of a cylindrical 
null dust collapse, originally constructed by Morgan \cite{Morgan:1973}. 
The feature of the singularity formed by the gravitational collapse of 
this null dust were studied by Letelier and Wang\cite{Letelier:1994}. 
They showed that in the case with only ingoing null dust, the singularity 
is the $p.p.$ curvature singularity which is defined by the behavior of 
components of the Riemann tensor with respect to {\it parallelly propagated} 
tetrad frame, while in the case of the coexistence of ingoing and outgoing 
null dust, the singularity is the {\it scalar polynomial} ($s.p.$) 
curvature singularity which is defined by the behavior of the scalar 
polynomials constructed by the Riemann tensor\cite{Hawking-Ellis;1973}. 
The global structure of the spacetime with only ingoing null dust 
has been studied  by Nolan\cite{Nolan:2002}. 

In this paper, we study the feature of the Riemann tensor and geodesics 
near the singularity in Morgan's spacetime from different point of view 
of Letelier and Wang. The assumption of only ingoing null dust corresponds 
to the boundary condition in which all the null dust is absorbed by 
the singularity formed by itself. However, of course, this boundary condition 
is not necessarily unique since, as mentioned, we do not know how to impose 
the boundary condition at the naked singularity. Keeping this fact in mind, 
there is a possibility that something is emitted from the naked singularity 
formed by the collapsed null dust although we do not know what something 
is at present. In this paper, we assume that the emitted matter or radiation 
is also null dust and study how the feature of the naked singularity changes 
by the amount of the emitted null dust. 

The paper is organized as follows. 
Section \ref{section:null} is devoted to a review of the cylindrical null 
dust collapse. In this section, in order to see the curvature 
strength of the naked singularity, we also investigate the components of 
the Riemann tensor. 
In section \ref{section:extension}, we investigate radial geodesics and their 
extendibility across the singularity. 
Section \ref{section:summary} summarizes the paper.
Appendix A gives a description of geodesics in the case 
of the Morgan's spacetime, which is used in section \ref{section:null}.
In Appendix B, we show the behavior of non-radial geodesics 
near the singularity. 
In Appendix C, we show the convergence of the Weyl tensor along null geodesics 
hitting the naked singularity.

In this paper, we adopt $c=1$ unit and follow the convention of the 
Riemann and metric tensors in Ref.\cite{Hawking-Ellis;1973}. The Greek indices 
mean general coordinate basis components while the Latin indices will be used 
to denote specific coordinate basis components. 

\section{Naked singularity in cylindrical null dust collapse} 
\label{section:null}

Let us consider the spacetime with whole cylinder 
symmetry\cite{Melvin:1964,Melvin:1965} and further, as mentioned, we assume 
the null dust. Then the line element of this spacetime is given as 
\begin{eqnarray} 
ds^2 = e^{2\gamma}(-dt^2+dr^2)+r^2d\phi^2+dz^2, 
\label{metric} 
\end{eqnarray} 
where $\gamma$ is a function of $t$ and $r$. 
The energy-momentum tensor of null dust is 
\begin{eqnarray} 
T_{\mu\nu}=\frac{1}{\kappa r} \left( A k_{\mu} k_{\nu}
+ B \ell_{\mu} \ell_{\nu}  \right),
\label{EMtensor}
\end{eqnarray} 
where $\kappa$ is the gravitational constant and $k_{\mu}$ is an ingoing null vector, 
$k_{\mu}=(-1,-1,0,0)$, while $\ell_{\mu}$ is an outgoing null vector,  
$\ell_{\mu}=(-1,1,0,0)$. 
The strong, weak and dominant energy conditions are all 
equivalent to $A\ge 0$ and $B\ge 0$. In the following, we assume these conditions.

Hereafter we shall also use null coordinates,
\begin{eqnarray}
u&=&t-r, \\
v&=&t+r. 
\end{eqnarray}
The local conservation law ${T_{\mu}^{~\nu}}_{;\nu}=0$ leads $(\partial_uA)_{v}=0$ 
and $(\partial_vB)_{u}=0$, and thus, $A=A(v)$ and $B=B(u)$.
We assume appropriate differentiability more than $C^1$ for $A(x)$ and 
$B(x)$ in $-\infty<x<+\infty$. 
In this case, the Einstein equations demand that $A$ and $B$ are expressed in terms 
of the metric function $\gamma$ as 
\begin{eqnarray}
A=(\partial_v\gamma)_{u}~, \qquad B=-(\partial_u\gamma)_{v}. \label{density}
\end{eqnarray}
The above equations and the local conservation law demand 
$\partial_u\partial_v\gamma=0$ which is consistent with the Einstein equations. 
The above equations also demand that $\gamma(x,y)$ is at least $C^2$ function defined 
in $-\infty<x<+\infty$ and $-\infty<y<+\infty$. 
The solution for $\gamma$ is then written as
\begin{equation}
\gamma=\alpha(v)+\beta(u), \label{eq:alpha-beta}
\end{equation}
and therefore we obtain
\begin{equation}
A(v)=\alpha_{,v}, \qquad B(u)=-\beta_{,u}, \label{einstein-eq}
\end{equation}
where a comma means the ordinary derivative with respect to the variable following it. 

It is seen from eq.(\ref{EMtensor}) that if $A$ or $B$ do not vanish at $r=0$, the 
coordinate basis components of the energy-momentum tensor diverge at $r=0$. This 
implies that the spacetime singularity appears at the symmetry axis $r=0$. Indeed, 
Leterier and Wang showed that the curvature singularity appears 
there \cite{Letelier:1994}. As mentioned in the previous section, the singularity 
is $p.p.$ curvature one in the case with only one of the ingoing null dust and 
outgoing one, while the singularity is $s.p.$ curvature singularity if the 
ingoing null dust coexists with the outgoing one at the symmetry axis. 
This singularity is, of course, naked, i.e., in principle, observable. 

In this paper, we assume that the null dust takes the form of 
a cylindrical shell with finite thickness. This is equivalent 
to the assumption of the compact supports of $A$ and $B$. 
Further we assume the situation in which before the ingoing null 
dust collapses to form the singularity at the symmetry axis $r=0$, 
there is no spacetime singularity. 
Morgan showed that a conical singularity forms at the symmetry axis 
after all of the null dust collapses, in the case of no outgoing null dust, 
i.e., $B=0$ \cite{Morgan:1973}.
However, as mentioned in the previous section, the law of 
physics at the spacetime singularity has not yet been known. Something 
might be emitted from the naked singularity and here we assume that 
this something is also the null dust and investigate the back-reaction 
effect on the naked singularity by the emission of the null dust. 

We investigate the curvature strength of the naked singularity. 
For this purpose, we consider an observer freely falling radially 
to the naked singularity, where `radially' means vanishing 
$\phi$-component of the 4-velocity vector of the observer. 
The coordinate basis components of the 4-velocity 
vector of this observer are given by
\begin{eqnarray}
\frac{dt}{d\tau}&=&\frac{(1+V_{\rm (r)})e^{-2\alpha}+(1-V_{\rm (r)})e^{-2\beta}}
{2\sqrt{1-V^2}}, 
\label{dtdtau} \\
\frac{dr}{d\tau}&=&\frac{(1+V_{\rm (r)})e^{-2\alpha}-(1-V_{\rm (r)})e^{-2\beta}}
{2\sqrt{1-V^2}},
\label{drdtau}  \\
\frac{dz}{d\tau}&=&\frac{V_{\rm (z)}}{\sqrt{1-V^2}}, 
 \label{dzdtau}
 \end{eqnarray}
where $\tau$ is the proper time of the observer, and 
$V_{\rm (r)}$ and $V_{\rm (z)}$ are arbitrary constants, 
and $V^2=V_{\rm (r)}^2+V_{\rm (z)}^2$. Note that $V^2$ should be smaller than unity. 
The derivation of the above solutions is given in Appendix A. 
Then we find the behavior of the Ricci tensor in approaching 
to the naked singularity as 
\begin{eqnarray}
\lim_{\tau\to0} \tau R_{\mu\nu}\frac{dx^{\mu}}{d\tau}\frac{dx^{\nu}}{d\tau}
&=&\lim_{\tau\to 0} \frac{\tau}{r} \cdot 
\frac{(1+V_{\rm (r)})^2\alpha_{,v}e^{-4\alpha}
-(1-V_{\rm (r)})^2\beta_{,u}e^{-4\beta}}{1-V^2} \nonumber \\
&=& \left[\lim_{\tau\to 0} \frac{d\tau}{dr}\right] 
\left[\lim_{\tau \to 0}
\frac{(1+V_{\rm (r)})^2Ae^{-4\alpha}
+(1-V_{\rm (r)})^2Be^{-4\beta}}{1-V^2}
\right] \nonumber \\
&=&\frac{2}{\sqrt{1-V^2}}\cdot
\frac{(1+V_{\rm (r)})^2Ae^{-4\alpha}+(1-V_{\rm (r)})^2Be^{-4\beta}}
{(1+V_{\rm (r)})e^{-2\alpha}-(1-V_{\rm (r)})e^{-2\beta}}\Biggr|_{r=0},
\end{eqnarray}
where the proper time $\tau$ is adjusted as it equals zero at $r=0$. 
Equation (\ref{einstein-eq}) and the l'Hospital theorem has been used in the second line. 
The final line also shows that the above limit does not 
vanish in the region filled with the null dust since $A\geq0$ and $B\geq0$. 
This is a sufficient condition for the strong curvature condition by Kr\'{o}lak, 
which is often called the limiting focusing condition (LFC) \cite{Krolak:1987,Clarke:1993}.  
In this sense, this naked singularity is strong.

However, this singularity satisfies neither the strong limiting forcusing condition (SLFC) 
nor the strong curvature condition by Tipler (SCST). 
We say that SLFC is satified if the following integral defined on
a future directed null geodesic terminating at the singularity is unbounded: 
\begin{eqnarray}
\lim_{\lambda\to 0} \int^{\lambda}_{\lambda_0}d\lambda' 
\int^{\lambda'}_{\lambda_0}d\lambda'' \ 
R_{\mu\nu}\frac{dx^{\mu}}{d\tau}\frac{dx^{\nu}}{d\tau},
\label{eq:int-SLFC}
\end{eqnarray}
where the affine parameter $\lambda$ is adjusted as it equals zero at $r=0$ and 
$\lambda_0$ is some negative value.
The components of the tangent vector for the radial null geodesic are
\begin{eqnarray}
\frac{dt}{d\lambda}&=&{1\over2}\left[(\omega+k_{\rm (r)})e^{-2\alpha(v)}
+(\omega-k_{\rm (r)}) e^{-2\beta(u)}\right], \\
\frac{dr}{d\lambda}&=&{1\over2}\left[(\omega+k_{\rm (r)})e^{-2\alpha(v)}
-(\omega-k_{\rm (r)})e^{-2\beta(u)}\right], \\
\frac{dz}{d\lambda}&=&k_{\rm (z)},
\end{eqnarray}
where $\omega$, $k_{\rm (r)}$ and $k_{\rm (z)}$ are integration constants 
which satisfy $\omega^2=k_{\rm (r)}^2+k_{\rm (z)}^2$.
These solutions are derived in Appendix A.
Then, the Ricci tensor behaves along the null geodesic as
\begin{eqnarray}
\lim_{\lambda\to0} \lambda R_{\mu\nu}\frac{dx^{\mu}}{d\lambda}\frac{dx^{\nu}}{d\lambda}
&=& 2 \cdot \frac{(\omega+k_{(r)})^2e^{-4\alpha}A+(\omega-k_{(r)})^2e^{-4\beta}B}
{(\omega+k_{(r)})e^{-2\alpha}-(\omega-k_{(r)})e^{-2\beta}}
\Biggr|_{r=0}  =: X.
\end{eqnarray}
Thus, the above limit does not vanish as in the case of timelike geodesics. 
The limiting value is named $X$.
In order to show that the singularity does not satisfy the SLFC, let us consider the following limit:
\begin{eqnarray}
\lim_{\lambda\to 0}\frac{\int^{\lambda}_{\lambda_0}d\lambda'\int^{\lambda'}_{\lambda_0}d\lambda''\ 
R_{\mu\nu}\frac{dx^{\mu}}{d\lambda}\frac{dx^{\nu}}{d\lambda} }
{\lambda\ln\lambda-\lambda} 
=\lim_{\lambda\to 0} \frac{\int^{\lambda}_{\lambda_0}d\lambda''\ 
R_{\mu\nu}\frac{dx^{\mu}}{d\lambda}\frac{dx^{\nu}}{d\lambda} }{\ln \lambda}=X,
\end{eqnarray}
where the l'Hospital theorem has been used.
Therefore, 
\begin{eqnarray}
\lim_{\lambda\to 0} \int^{\lambda}_{\lambda_0}d\lambda' \int^{\lambda'}_{\lambda_0}
d\lambda''\ 
R_{\mu\nu}\frac{dx^{\mu}}{d\lambda}\frac{dx^{\nu}}{d\lambda} =\lim_{\lambda\to 0}\Big(
X(\lambda\ln\lambda-\lambda)
\Big)=0.
\end{eqnarray}
Thus, the integral (\ref{eq:int-SLFC}) is integrable and the strong limiting forcusing condition is
not satisfied.
Additionally, the integral for the Weyl tensors 
\begin{eqnarray}
L^{a}_{b}=\lim_{\lambda\to0}\int^{\lambda}_{\lambda_0}d\lambda'\int^{\lambda'}_{\lambda_0}d\lambda''
\left(\int^{\lambda''}_{\lambda_0}d\lambda'''\biggr|C^{a}_{\ \mu b\nu}
\frac{dx^{\mu}}{d\lambda}\frac{dx^{\nu}}{d\lambda}
\biggl| \right)^2
\label{eq:int-Weyl}
\end{eqnarray}
converges for all $a$ and $b$ as shown in Appendix \ref{sec:Weyl}. 
And now, a necessary condition for SCST is
that either the integral (\ref{eq:int-SLFC}) or (\ref{eq:int-Weyl}) does not converge \cite{Clarke:1993}.
Therefore, this singularity does not satisfy the necessary condition for SCST.

In the above argument, we discuss only for the null geodesic. 
As we show in Appendix \ref{sec:timelikeSCST},  
the singularity also does not satisfy the necessary condition in the case of timelike geodesic.
Thus, the singularity does not satisfy SCST.

\section{Geodesics in the neighborhood of the singularity}
\label{section:extension}

Here we focus on the radial null and timelike geodesics which necessarily hit 
the singularity at $r=0$. 
In order to see the behavior of those near the singularity, let us adopt 
a Cartesian coordinate instead of the cylindrical coordinate (\ref{metric}),
which is defined by
\begin{eqnarray}
x:=r \cos \phi,\qquad y:= r \sin \phi.
\end{eqnarray}
Then, the components of the metric tensor with respect to the new coordinate basis 
are given as
\begin{eqnarray}
g_{xx} &=& \frac{x^2 e^{2\gamma}+y^2}{r^2} = e^{2\gamma}\cos^2\phi+\sin^2\phi, \\
g_{xy} &=& \frac{xy(e^{2\gamma}-1)}{r^2} =(e^{2\gamma}-1)\sin\phi\cos\phi, \\
g_{yy} &=& \frac{x^2+e^{2\gamma}y^2}{r^2}=\cos^2\phi+e^{2\gamma}\sin^2\phi. 
\end{eqnarray}
The Christoffel symbols in this coordinate system are given as follows: 
\begin{eqnarray}
\Gamma^t_{tt}&=&\dot{\gamma},~~~~~~~~~~~\Gamma^t_{tx}=\gamma'\cos\phi,~~~~~~~~~
\Gamma^t_{ty}=\gamma'\sin\phi, \\
\Gamma^t_{xx}&=&\dot{\gamma}\cos^2\phi,~~~
\Gamma^t_{xy}=\dot{\gamma}\sin\phi\cos\phi,~~~
\Gamma^t_{yy}=\dot{\gamma}\sin^2\phi, \\
\Gamma^x_{tt}&=&\gamma'\cos\phi,~~~
\Gamma^x_{tx}=\dot{\gamma}\cos^2\phi, ~~~~~~~~~
\Gamma^x_{ty}=\dot{\gamma}\sin\phi\cos\phi, \\
\Gamma^x_{xx}&=&\cos\phi\left(\gamma'\cos^2\phi
+\frac{1-e^{-2\gamma}}{r}\sin^2\phi\right), \\
\Gamma^x_{xy}&=&\sin\phi\cos^2\phi\left(\gamma'-\frac{1-e^{-2\gamma}}{r}\right), \\
\Gamma^x_{yy}&=&\cos\phi\left(\gamma'\sin^2\phi
+\frac{1-e^{-2\gamma}}{r}\cos^2\phi\right), \\
\Gamma^y_{tt}&=&\gamma'\sin\phi,~~~
\Gamma^y_{tx}=\dot{\gamma}\sin\phi\cos\phi ,~~~~~
\Gamma^y_{ty}=\dot{\gamma}\sin^2\phi, \\
\Gamma^y_{xx}&=&\sin\phi\left(\gamma'\cos^2\phi
+\frac{1-e^{-2\gamma}}{r}\sin^2\phi\right), \\
\Gamma^y_{xy}&=&\sin^2\phi\cos\phi\left(\gamma'-\frac{1-e^{-2\gamma}}{r}\right),\\
\Gamma^y_{yy}&=&\sin\phi\left(\gamma'\sin^2\phi
+\frac{1-e^{-2\gamma}}{r}\cos^2\phi\right), 
\end{eqnarray}
and the other components vanish, where $\dot{\gamma}=(\partial_t\gamma)_r$ and 
$\gamma'=(\partial_r\gamma)_t$. 

By virtue of the cylindrical symmetry, it is sufficient for our purpose to consider 
only the case in which the initial data is given by $x<0$, $y=0$, $dx/d\tau>0$ and 
$dy/d\tau=0$ at some proper time $\tau$. We can easily see that $y$ remains 
zero over the singularity $r=0$ since $\Gamma^{y}_{\mu\nu}=0$ on $y=0$.  
Due to the translational symmetry of $z$-direction, 
$dz/d\tau$ is constant and is denoted by the constant $V_{\rm (z)}$ as in 
the previous section. Then the remaining geodesic equations are given by
\begin{eqnarray}
\frac{d^2t}{d\tau^2}
&=&-\dot{\gamma}\left(\frac{dt}{d\tau}\right)^2
-2\frac{x}{|x|}\gamma' \frac{dt}{d\tau}\frac{dx}{d\tau}
-\dot{\gamma}\left(\frac{dx}{d\tau}\right)^2, \\
\frac{d^2x}{d\tau^2}
&=&-\frac{x}{|x|}\gamma'\left(\frac{dt}{d\tau}\right)^2
-2\dot{\gamma}\frac{dt}{d\tau}\frac{dx}{d\tau}
-\frac{x}{|x|}\gamma'\left(\frac{dx}{d\tau}\right)^2.
\end{eqnarray}
The right hand sides of the above equations are multi-valued at $x=0$ due to $x/|x|$. 
However those are finite and thus the solutions for $dt/d\tau$ and 
$dx/d\tau$ are $C^{1-}$ functions of $\tau$. This result seems to be likely  
that the geodesics which hit the singularity $r=0$ are uniquely extendible across the 
singularity, but it is no true except for the case of $\gamma|_{r=0}=0$. 

In order to see this fact, we explicitly write down 
the norm $N$ of the geodesic tangent at $r=0$ as
\begin{eqnarray}
N=\left[-e^{2\gamma}\left(\frac{dt}{d\tau}\right)^2
  +\left(e^{2\gamma}\cos^2\phi+\sin^2\phi\right)\left(\frac{dx}{d\tau}\right)^2
  \right]\Biggr|_{r=0}
  +V_{\rm (z)}^2,
\label{eq:norm}
\end{eqnarray}
where we should note that arbitrary values of $\phi$ are assigned for a point on the 
symmetry axis $r=0$ and thus $\phi$ appears in the right hand side 
of the above equation. Since $dt/d\tau$ and $dx/d\tau$ are continuous, it is 
seen from eq.(\ref{eq:norm}) that the norm $N$ becomes multi-valued  
at $r=0$ as long as $\gamma$ does not vanish at the symmetry axis $r=0$. 
Here it should be noted that the norm of the affinely parametrized geodesic 
tangent is preserved. 
Therefore geodesics which hit the naked singularity at $r=0$ with 
non-vanishing $\gamma$ can not be extended across this naked singularity 
since the statement that the geodesic tangent is parallelly propagated is 
meaningless there. Expressed in another way, the spacetime is conical around $r=0$. 
The reason of the incompleteness is the same as that of the fact that any spacetime 
with conical singularities is geodesically incomplete.

Here we consider a particular but physically important case in 
which $\gamma$ always vanishes at $r=0$ 
and thus $\dot{\gamma}$ should also always vanish there. 
Since $\dot{\gamma}=(\partial_v\gamma)_u+(\partial_u\gamma)_v=A-B$ 
by eq.(\ref{einstein-eq}), the condition $\dot{\gamma}|_{r=0}=0$ leads
\begin{eqnarray}
A(t)=B(t). \label{pass-condition}
\end{eqnarray}
The above equation means that the same amount of outgoing null dust 
as that of ingoing one is emitted from the naked singularity as soon as the ingoing null 
dust hits the symmetry axis and forms the naked singularity there. In this case, 
the components of the metric tensor with respect to the Cartesian coordinate 
system are single valued and continuous because at the naked singularity 
$r=0$, $g_{tt}=-1=-g_{xx}=-g_{yy}=-g_{zz}$ and the other components vanishes. 
Thus even at the singularity, the norm $N$ of the radial geodesics is 
uniquely determined and further it is distinguishable 
whether the two vectors are parallel with each other. 
This fact guarantees that radial geodesics can be extended 
uniquely across the singularity. Non-radial causal geodesics 
might be unable to hit the singularity in the present case (see Appendix B), 
and therefore the present spacetime will be geodesically complete even 
though there is a spacetime singularity which satisfies LFC. 

The completeness of radial geodesics means that the 
trajectories of the particles constituting the null dust are extendible and hence 
the condition (\ref{pass-condition}) implies that each particle of the 
null dust passes through the naked singularity at the symmetry axis $r=0$. 
Therefore, the `boundary condition' (\ref{pass-condition}) at the naked singularity 
may be called {\it passing-through condition}. 
Further it should be noted that after all of the collapsed null dust leaves 
the naked singularity in the form of the outgoing one, the singularity disappears 
and the flat spacetime remains. 

\section{Summary and discussion}
\label{section:summary}

We studied the gravitational collapse of cylindrically symmetric null dust 
to form a naked singularity at its symmetry axis. 
We assumed on the situation in which the null dust is emitted 
from the naked singularity formed by the collapsed null dust. 
Then we investigated the curvature strength of the naked singularity and 
the behavior of geodesics near this naked singularity to see the back-reaction 
effect by the emission of null dust from the naked singularity. 

We found that if the same amount of outgoing null dust 
as that of ingoing one is emitted from the naked singularity as soon as the 
ingoing null dust hits the symmetry axis and forms the naked singularity there, 
the spacetime will be geodesically complete, even if there is the naked 
singularity which satisfies LFC. 
The completeness of radial geodesics means that the 
trajectories of the particles constituting the null dust are uniquely extendible 
across the naked singularity and hence 
this case is regarded as the situation in which each particle of the 
null dust passes through the naked singularity and then leaves for 
infinite. Therefore, this boundary condition at the naked singularity 
has been named the passing-through condition. 
Further it should be noted that after all of the collapsed null dust 
leaves the naked singularity in the form of the outgoing null dust, the 
singularity disappears and the flat spacetime remains. 

Usually, LFC is thought as a quite strong condition for a singularity \cite{Clarke:1993}.
However in the case of passing-through condition, we can extend geodesics 
across the singularity and resolve it. Thus, this singularity seems to be not so 
strong and severe. It implies that LFC 
does not necessarily represent the strength of spacetime singularities. 

If all of the interactions becomes weaker in higher energy state due to the 
asymptotic freedom, the passing-through condition on the null dust seems to be 
natural. Thus null dust collapse with passing-through condition 
might be a toy model describing the essence of a ``realistic naked 
singularity formation''.

\begin{acknowledgments}
It is our pleasure to thank Hideki Ishihara for his valuable discussion 
and also to thank Hideo Kodama for his important comment on the geodesically 
completeness. We are grateful to colleagues in the astrophysics and 
gravity group of Osaka City University for helpful discussion and criticism. 
This work is supported by the Grant in Aid for Scientific Research (No.16540264) 
from JSPS. YK thanks the Yukawa Memorial Foundation for its support.  
\end{acknowledgments}

\appendix
\section{Radial geodesics}
\label{section:geodesics}

We derive the tangent vectors of radial geodesics in Morgan spacetime discussed 
in section \ref{section:null}. 
Here we adopt the null coordinate $u$ and $v$ and thus the Lagrangian of a 
geodesic is given by
\begin{eqnarray}
L=-e^{2\gamma}\frac{du}{d\lambda}\frac{dv}{d\lambda}+\left(\frac{dz}{d\lambda}\right)^2,
\end{eqnarray}
where $\lambda$ is the affine parameter. 
The geodesic equations for the radial geodesic are
\begin{eqnarray}
\frac{d^2v}{d\lambda^2}+2(\partial_v\gamma)_{u}\left(\frac{dv}{d\lambda}\right)^2&=&0, \\
\frac{d^2u}{d\lambda^2}+2(\partial_u\gamma)_{v}\left(\frac{du}{d\lambda}\right)^2&=&0, \\
\frac{d^2z}{d\lambda^2}&=&0. \label{z-eq}
\end{eqnarray}
Substituting eq.(\ref{eq:alpha-beta}) into the above equation, we obtain
\begin{eqnarray}
\frac{d}{d\lambda}\left(e^{2\alpha}\frac{dv}{d\lambda}\right)=0,
\qquad 
\frac{d}{d\lambda}\left(e^{2\beta}\frac{du}{d\lambda}\right)=0. 
\end{eqnarray}
Since the integrations of the above equations and eq.(\ref{z-eq}) 
are trivial, we obtain 
\begin{eqnarray}
e^{2\alpha}\frac{dv}{d\lambda}={\rm const.},
\qquad 
e^{2\beta}\frac{du}{d\lambda}={\rm const}~~~~~~{\rm and}~~~~~~\frac{dz}{d\lambda}={\rm const}.
\end{eqnarray}
Therefore for the null geodesics, we get
\begin{eqnarray}
\frac{dt}{d\lambda}&=&{1\over2}\left[(\omega+k_{\rm (r)})e^{-2\alpha(v)}
+(\omega-k_{\rm (r)})
e^{-2\beta(u)}\right], 
\label{eq:null-geodesic-t}  \\
\frac{dr}{d\lambda}&=&{1\over2}\left[(\omega+k_{\rm (r)})e^{-2\alpha(v)}
-(\omega-k_{\rm (r)})e^{-2\beta(u)}\right], 
\label{eq:null-geodesic-r} \\
\frac{dz}{d\lambda}&=&k_{\rm (z)},
\label{eq:null-geodesic-z}
\end{eqnarray}
where $\omega$, $k_{\rm (r)}$ and $k_{\rm (z)}$ are integration constants 
which satisfy $\omega^2=k_{\rm (r)}^2+k_{\rm (z)}^2$. Note that in the case of 
$\alpha=\beta=0$, the components of the tangent vector become 
\begin{equation}
\frac{dt}{d\lambda}=\omega,~~~
\frac{dr}{d\lambda}=k_{\rm (r)}~~~{\rm and}~~~
\frac{dz}{d\lambda}=k_{\rm (z)}.
\end{equation}
On the other hand, for the timelike geodesics, we obtain
\begin{eqnarray}
\frac{dt}{d\lambda}&=&\frac{(1+V_{\rm (r)})e^{-2\alpha}+(1-V_{\rm (r)})e^{-2\beta}}
{2\sqrt{1-V^2}}, \\
\frac{dr}{d\lambda}&=&\frac{(1+V_{\rm (r)})e^{-2\alpha}-(1-V_{\rm (r)})e^{-2\beta}}
{2\sqrt{1-V^2}}, \\
\frac{dz}{d\lambda}&=&\frac{V_{\rm (z)}}{\sqrt{1-V^2}}, 
 \end{eqnarray}
where $V_{\rm (r)}$ and $V_{\rm (z)}$ are integration constants and 
$V^2=V_{\rm (r)}^2+V_{\rm (z)}^2$ should be smaller than unity. 
Note that in the case of $\alpha=\beta=0$, 
the components of the tangent vector become
\begin{eqnarray}
\frac{dt}{d\lambda}=\frac{1}{\sqrt{1-V^2}},~~~
\frac{dr}{d\lambda}=\frac{V_{\rm (r)}}{\sqrt{1-V^2}}~~~{\rm and}~~~
\frac{dz}{d\lambda}=\frac{V_{\rm (z)}}{\sqrt{1-V^2}}.
\end{eqnarray}

\section{Non-Radial Geodesics near the singularity}
\label{section:not-hit}

We investigate the behavior of non-radial geodesics near the naked singularity 
in the case of $A(t)=B(t)$. Adopting the coordinate system (\ref{metric}), 
the geodesic equations are written as
\begin{eqnarray}
\frac{d}{d\lambda}\left(e^{2\gamma}\frac{dt}{d\lambda}\right)&=&
-\dot{\gamma}e^{2\gamma}\left[-\left(\frac{dt}{d\lambda}\right)^2
+\left(\frac{dr}{d\lambda}\right)^2\right], \label{1st-eq}\\
\frac{d}{d\lambda}\left(e^{2\gamma}\frac{dr}{d\lambda}\right)&=&
\gamma'e^{2\gamma}\left[-\left(\frac{dt}{d\lambda}\right)^2
+\left(\frac{dr}{d\lambda}\right)^2\right]+r\left(\frac{d\phi}{d\lambda}\right)^2, \\
\frac{d}{d\lambda}\left(r^2\frac{d\phi}{d\lambda}\right)&=&0, \\
\frac{d^2z}{d\lambda^2}&=&0, 
\end{eqnarray}
where $\lambda$ is the affine parameter. 
The third and forth equations can be easily integrated once and we obtain
\begin{equation}
r^2\frac{d\phi}{d\lambda}=L, \qquad \frac{dz}{d\lambda}=P,
\end{equation}
where $L$ and $P$ are integration constants. 
The non-radial geodesic means the geodesic with non-vanishing $L$. 
Then the normalization of the geodesic tangent vector leads
\begin{eqnarray}
\left(\frac{dr}{d\lambda}\right)^2
=\left(\frac{dt}{d\lambda}\right)^2+e^{-2\gamma}\left(\chi-\frac{L^2}{r^2}-P^2\right), 
\label{normalization}
\end{eqnarray}
where $\chi$ is equal to $-1$ for timelike geodesics, unity for spacelike geodesics 
and vanishes for null geodesics. 
Since the left hand side of the above equation is non-negative, we obtain an 
inequality as
\begin{equation}
r^2\geq \frac{L^2}{e^{2\gamma}\left(dt/d\lambda\right)^2-(P^2-\chi)}.
\end{equation}
Hence in order that the geodesic of non-vanishing $L$ hits the naked singularity at $r=0$, 
$dt/d\lambda$ should become infinite in approaching 
to the naked singularity. 

Here we adjust the affine parameter $\lambda$ and coordinate time $t$ 
so that the geodesic hits the singularity at $\lambda=0$ and $t=0$. 
Thus, the geodesic which hits the naked singularity should behave 
near the naked singularity as
\begin{eqnarray}
t\sim C_{\rm (t)}\lambda^p~~~~~{\rm and}~~~~~ r\sim C_{\rm (r)}\lambda^q,
\label{behavior}
\end{eqnarray}
where $C_{\rm (t)}$ and $C_{\rm (r)}$ are constant, 
and $0<p<1$ should be satisfied since $dt/d\lambda\to \infty$ for $\lambda\to\infty$.

Hereafter our purpose is to show that there is no solution consistent 
with the asymptotic behavior eq.(\ref{behavior}). 
From eq.(\ref{normalization}), we find the following behavior  
in the limit $\lambda\to 0$, or equivalently, $r\to 0$,  
\begin{equation}
\left(\frac{dr}{d\lambda}\right)^2
\sim
\left(\frac{dt}{d\lambda}\right)^2-e^{-2\gamma_0}\frac{L^2}{r^2},
\end{equation}
where $\gamma_0$ is the value of $\gamma$ when the geodesic hits the naked singularity.  
Of course, the following inequality should be satisfied:
\begin{equation}
\left(\frac{dt}{d\lambda}\right)^2>e^{-2\gamma_0}\frac{L^2}{r^2}~~~{\rm for}~~\lambda\longrightarrow0.
\end{equation}
Therefore $q$ should be equal to $p$ and further from the above inequality, we 
obtain $p\leq1/2$, because the left hand side is proportional to $\lambda^{2(p-1)}$ while 
the right hand side is proportional to $\lambda^{-2q}=\lambda^{-2p}$.  

By using the normalization condition of the tangent vector, 
eq.(\ref{1st-eq}) is rewritten as
\begin{equation}
\frac{d^2t}{d\lambda^2}+2\dot{\gamma}\left(\frac{dt}{d\lambda}\right)^2
+2\gamma'\left(\frac{dt}{d\lambda}\right)\left(\frac{dr}{d\lambda}\right)
-\dot{\gamma}e^{-2\gamma}\left(P^2+\frac{L^2}{r^2}-\chi \right)=0.
\label{t-eq}
\end{equation}
We consider the case in which $\gamma'$, $\ddot{\gamma}$ and $\dot{\gamma}'$ 
are finite even at the naked singularity 
$r=0$ and their values at the naked singularity are denoted by 
$\gamma'_0$, $\ddot{\gamma}_0$ and $\dot{\gamma}'_0$, respectively. 
Note that $\dot{\gamma}$ vanishes at $r=0$ by the assumption $A(t)=B(t)$ 
and thus it behaves near the naked singularity as
\begin{equation}
\dot{\gamma} \sim 
\left(\ddot{\gamma}_0\frac{dt}{d\lambda}+\dot{\gamma}'_0
\frac{dr}{d\lambda}\right)\lambda \sim 
\left(\ddot{\gamma}_0C_{\rm (t)}+\dot{\gamma}'_0C_{\rm (r)}\right)p\lambda^p.
\end{equation}
Keep the above behavior in mind and substituting the asymptotic behavior 
eq.(\ref{behavior}) into eq.(\ref{t-eq}), we obtain 
\begin{eqnarray}
&&\left[ {\rm The}~~{\rm left}~~{\rm hand}~~{\rm side}~{\rm of}
~~{\rm eq.(\ref{t-eq})} \right] \nonumber \\
&&~~~~\sim~~ p(p-1)C_{\rm (t)}\lambda^{p-2}
+ 2\left(\ddot{\gamma}_0C_{\rm (t)}+\dot{\gamma}'_0C_{\rm (r)}\right)
C_{\rm (t)}p^2\lambda^{3p-2}
+2\gamma_0'C_{\rm (t)} C_{\rm (r)}p^2\lambda^{2(p-1)}  \nonumber \\ 
&& ~~~~~~~~~~~~
-\left(\ddot{\gamma}_0C_{\rm (t)}+\dot{\gamma}'_0C_{\rm (r)}\right)
pe^{-2\gamma_0}\frac{L^2}{C_{\rm (r)}^2}\lambda^{-p}.
\end{eqnarray}
The leading order terms should be balanced with each other. 
One can easily see that the second term can not be one of the leading order terms.
If the 2nd term is so, $3p-2$ should be less than or equal to each of $p-2,~ 2(p-1)$ 
and $(-p)$, because we consider the limit $\lambda \to 0$.
This leads $p\le 0$, conflicting with $p>0$. 
Thus this is impossible.  
If the 1st term is balanced with the 3rd term, then $p-2=2(p-1)$. This leads $p=0$ and 
thus it is impossible because $p>0$. 
If the 1st term is balanced with the last term, then 
$p-2=-p$ and this leads $p=1$. This conflicts with $p\leq 1/2$ and thus 
this is impossible. Finally, if the 3rd term is balanced with the last term, 
$2(p-1)=-p$ and this leads $p=2/3$. This also conflicts with $p \leq 1/2$. 
Therefore there is no solution which behaves as eq.(\ref{behavior}). This 
fact strongly suggests that there is no non-radial geodesic which hits the 
naked singularity at $r=0$.

\section{Weyl tensor}
\label{sec:Weyl}

We investigate the behavior of Weyl tensor in approaching to the neked singularity.
The components of Weyl tensor in the coordinate system (\ref{metric}) are
\begin{eqnarray}
&&C_{t\phi t\phi}=C_{r\phi r\phi}=-r^2 C_{tztz}=-r^2C_{rzrz}=\frac{1}{2}r \gamma' \\
&&C_{t\phi t\phi}=-r^2 C_{tzrz}=\frac{1}{2}r \dot{\gamma},
\end{eqnarray}
and the other components vanish, where $\dot{\gamma}=(\partial_t \gamma)_{r}$ and $\gamma'=(\partial_{r} \gamma)_{t}$.
In order to investigate their behavior along the null geodesic defined by (\ref{eq:null-geodesic-t}), 
(\ref{eq:null-geodesic-r}) and (\ref{eq:null-geodesic-z}), 
we intorduce a parallely propagating frame as follows:
\begin{eqnarray}
e_{(\lambda)}^{\mu} &=& \left(\frac{(\omega+k_{(r)})e^{-2\alpha}+(\omega-k_{(r)})e^{-2\beta}}{2}, 
\frac{(\omega+k_{(r)})e^{-2\alpha}-(\omega-k_{(r)})e^{-2\beta}}{2}, 0, k_{(z)}\right)_{tr\phi z} \\
e_{(\ell)}^{\mu} &=& \left(\frac{e^{-2\alpha}+e^{-2\beta}}{2(\omega-k_{(z)})}, 
\frac{e^{-2\alpha}-e^{-2\beta}}{2(\omega-k_{(z)})}, 0, \frac{1}{\omega-k_{(z)}}\right)_{tr\phi z} \\
e_{(\phi)}^{\mu}&=&\left(0,0,\frac{1}{r}, 0\right)_{tr\phi z}, \\
e_{(s)}^{\mu} &=& \bigg(\frac{c_{\alpha}e^{-2\alpha}-c_{\beta}e^{-2\beta}}{2(\omega-k_{(z)})}, 
\frac{c_{\alpha}e^{-2\alpha}+c_{\beta}e^{-2\beta}}{2(\omega-k_{(z)})}, 0, 
\frac{k_{(r)}}{\omega-k_{(z)}}\bigg)_{tr\phi z} 
\end{eqnarray} 
where we set $c_{\alpha}=\omega+k_{(r)}-k_{(z)}, c_{\beta}=\omega-k_{(r)}-k_{(z)}$.
These vectors satisfy the following relations:
\begin{eqnarray}
&&e_{(a)}^{\mu} e_{(a)\mu} = 0, \quad e_{(\lambda)}^{\mu} e_{(\ell)\mu} = -1, \quad (a=\lambda, \ell ), \\
&&e_{(b)}^{\mu} e_{(b)\mu} = 1, \quad e_{(s)}^{\mu} e_{(\phi)\mu}=0,\quad 
e_{(a)}^{\mu}e_{(b)\mu} =0, \quad (b=s,\phi)  \\
&&e_{(a)\mu;\nu}e_{(\lambda)}^{\nu} = e_{(b)\mu;\nu}e_{(\lambda)}^{\nu} = 0.
\end{eqnarray}
In this frame, the non-vanishing components of the Weyl tensor 
in the form of $C_{(\lambda)(*)(\lambda)(*)}$ are
\begin{eqnarray}
C_{(\lambda)(\phi)(\lambda)(\phi)} &=& 
\frac{\alpha_{,v}e^{-4\alpha}(\omega+k_{(r)})^2-\beta_{,u}e^{-4\beta}(\omega-k_{(r)})^2}{2r}, \\ 
C_{(\lambda)(s)(\lambda)(s)} &=& 
\frac{-\alpha_{,v}e^{-4\alpha}(\omega+k_{(r)})^2+\beta_{,u}e^{-4\beta}(\omega-k_{(r)})^2}{2r}.
\end{eqnarray}
Thus, all the above components have the following behavior:
\begin{eqnarray}
C_{(\lambda)(*)(\lambda)(*)} = \frac{f(\lambda)}{r},\quad  
\end{eqnarray}
where $f(\lambda)$ is a function having non-zero finite value at $\lambda=0$. 
We set the following limiting value as $Y$:
\begin{eqnarray}
Y&:=&\lim_{\lambda\to 0} \frac{\int^{\lambda}_{\lambda_0} d\lambda' \Big|f(\lambda')/r \Big|}{\log \lambda}
=\lim_{\lambda\to 0} \frac{\lambda}{r} |f(\lambda)|=\lim_{\lambda\to 0} \frac{d\lambda}{dr} |f(0)| \\
&=& \frac{2|f(0)|}{(\omega+k_r)e^{-2\alpha}-(\omega-k_r)e^{-2\beta}\Big|_{\lambda=0} },
\end{eqnarray}
where $\lambda_0$ is some negative value.
In order to show the convergence of the limit (\ref{eq:int-Weyl}), let us consider the following limit:
\begin{eqnarray}
\lim_{\lambda\to0} \frac{\int^{\lambda}_{\lambda_0}d\lambda' \int^{\lambda'}_{\lambda_0}d\lambda''
\left(\int^{\lambda''}_{\lambda_0}d\lambda'''\Big| f(\lambda''')/r \Big| \right)^2 }
{\frac{1}{2}\lambda^2(\log \lambda)^2-\frac{3}{2}\lambda^2 \log \lambda +\frac{7}{4}\lambda^2}
=\lim_{\lambda\to0 } \frac{\int^{\lambda}_{\lambda_0}d\lambda'\left(\int^{\lambda'}_{\lambda_0}d\lambda''
|f(\lambda'')/r|\right)^2 }{\lambda(\log \lambda)^2-2\lambda\log \lambda +2\lambda} \\
=\lim_{\lambda\to 0} \frac{\left( \int^{\lambda}_{\lambda_0}d\lambda'\Big|f(\lambda')/r \Big| 
\right)^2}{(\log \lambda)^2}=Y^2, 
\end{eqnarray}
where we have used the l'Hospital theorem.
Therefore, we have
\begin{eqnarray}
\lim_{\lambda\to0}
\int^{\lambda}_{\lambda_0}d\lambda' \int^{\lambda'}_{\lambda_0}d\lambda''
&&\left(\int^{\lambda''}_{\lambda_0}d\lambda'''\Big| f(\lambda)/r \Big| \right)^2 \\
&&\qquad =Y^2 \lim_{\lambda\to 0} \left(\frac{1}{2}\lambda^2(\log \lambda)^2-\frac{3}{2}\lambda^2 \log \lambda +\frac{7}{4}\lambda^2\right) \\
&&\qquad = 0.
\end{eqnarray}
Thus, for all non-vanishing Weyl components, the limit (\ref{eq:int-Weyl}) converges. 

\section{SCST for timelike geodesics}
\label{sec:timelikeSCST}

In this Appedix, we show that the naked singularity does not satisfy the necessary condition for SCST
in the case of timelike geodesics.
The necessary condition in this case is as follows: 
{\it for some component $R_{(\tau)(i)(\tau)(j)}$ of the Riemann tensor in a parallely propagated frame, the integral}
\begin{eqnarray}
I_{(i)(j)}(\tau) = \int^{\tau}_{\tau_0} d\tau' \int^{\tau'}_{\tau_0} d\tau'' \Big| R_{(\tau)(i)(\tau)(j)}(\tau'')  \Big|,
\end{eqnarray}
{\it does not converge as $\tau \to 0$, where the proper time $\tau$ is adjusted as it equals zero at $r=0$ and 
$\tau_0$ is some negative value. }
A parallely propagating frame along the timelike geodesic (\ref{dtdtau})-(\ref{dzdtau}) is 
\begin{eqnarray}
e_{(\tau)}^{\mu} &=& \left( \frac{dt}{d\tau}, \frac{dr}{d\tau}, 0 ,\frac{dz}{d\tau} \right)_{tr\phi z}, \\
e_{(n)}^{\mu} &=& \frac{1}{\sqrt{1+(dz/d\tau)^2}}  \left( \frac{dr}{d\tau}, \frac{dt}{d\tau},0,0  \right)_{tr\phi z}, \\
e_{(\phi)}^{\mu} &=& \left( 0,0, \frac{1}{r},0 \right)_{tr\phi z}, \\
e_{(u)}^{\mu} &=& 
\frac{1}{\sqrt{1+(dz/d\tau)^2}}  \left( \frac{dz}{d\tau}\frac{dr}{d\tau}, \ \frac{dz}{d\tau}\frac{dt}{d\tau}, \ 
0,\  1+\left(\frac{dz}{d\tau}\right)^2 \right)_{tr\phi z}.
\end{eqnarray}
The only non-vanishing component of the Rimann tensor in this frame is
\begin{eqnarray}
R_{(\tau)(\phi)(\tau)(\phi)} = \frac{1}{r(1-V^2)}\left[
(1+V_{(r)})^2 A e^{-4\alpha}+(1- V_{(r)})^2Be^{-4\beta}
\right].
\end{eqnarray}
We set the following limiting value as $Z$:
\begin{eqnarray}
Z:&=& \lim_{\tau \to 0}  \tau \bigg| R_{(\tau)(\phi)(\tau)(\phi)} \bigg| \\
&=& \lim_{\tau\to 0} \frac{\tau}{r}
\bigg| \frac{1}{(1-V^2)}\left[
(1+V_{(r)})^2 A e^{-4\alpha}+(1- V_{(r)})^2Be^{-4\beta}
\right] \bigg| \\
&=& \lim_{\tau\to 0} \frac{d\tau}{dr} \bigg| \frac{1}{(1-V^2)}\left[
(1+V_{(r)})^2 A e^{-4\alpha}+(1- V_{(r)})^2Be^{-4\beta}
\right] \bigg| \\
&=& \frac{2}{\sqrt{1-V^2}} \bigg| \frac{ (1+V_{(r)})^2Ae^{-4\alpha}+(1-V_{(r)})^2Be^{-4\beta}}
{(1+V_{(r)})e^{-2\alpha}+(1-V_{(r)})e^{-2\beta}}
\bigg|_{\tau =0},
\end{eqnarray}
where we have used the l'Hopital theorem.
Then,
\begin{eqnarray}
\lim_{\tau\to 0} \frac{\int^{\tau}_{\tau_0} d\tau' \int^{\tau'}_{\tau_0} d\tau'' \Big| R_{(\tau)(\phi)(\tau)(\phi)}(\tau'') 
\Big| }{\tau \ln \tau -\tau} = \lim_{\tau\to 0}\frac{\int^{\tau}_{\tau_0}
d\tau' \bigg|R_{(\tau)(\phi)(\tau)(\phi)}(\tau')\bigg|}{\ln\tau} \\
=\lim_{\tau\to 0}\tau \bigg|R_{(\tau)(\phi)(\tau)(\phi)}  (\tau) \bigg| = Z.
\end{eqnarray}
Thereofre, we have 
\begin{eqnarray}
I_{(i)(j)}(\tau) = \int^{\tau}_{\tau_0} d\tau' \int^{\tau'}_{\tau_0} d\tau'' \Big| R_{(\tau)(i)(\tau)(j)}(\tau'')  \Big|
=Z \lim_{\tau \to 0}\left( \tau \ln\tau -\tau\right) = 0.
\end{eqnarray}
The singularity does not satisfy the necessary condition for SCST.


\end{document}